\documentstyle[11pt,IAU207_pasp,twoside,psfig]{article}
\def\cent      {$\omega$\thinspace Centauri}
\def\tuca      {47\thinspace Tucanae}
\def\cen       {$\omega$\thinspace Cen}

\def\arcs               {$^{\prime\prime}$} 
\def\conc               {$c$ = log ($r_t/r_c$)}
\def\etal      {et\thinspace al.}
\def\ho        {$H_\circ$}
\def\kms       {km\thinspace s$^{-1}$}
\def\Mass               {$\mathcal{M}$}
\def\masyr     {mas\thinspace yr$^{-1}$}
\def\milm      {$10^6$$\mathcal{M}$$_{\odot}$}
\def\mily               {${ {10^6}}\thinspace $yr}
\def\msun      {$\mathcal{M}$$_{\odot}$}
\def\ml       {$\mathcal{M}$$/L$}
\def\mlv       {$\mathcal{M}$$/L_V$}
\def\Mtot      {$\mathcal{M}$$_{tot}$}

\def\muov               {$\mu$(0,$V$)}
\def\Mv                 {$M_V$}
\def\pmm       {$\pm$}
\def\rc        {$r_c$}

\def\rt        {$r_t$}
\def\sobs               {$\sigma_{obs}$}
\def\sigpo              {$\sigma_p$(0)}

\def\edcomment#1{\iffalse\marginpar{\raggedright\sl#1\/}\else\relax\fi}
\reversemarginpar

\begin{document}
\title{Mass Determinations of Star Clusters}
\author{Georges Meylan}
\affil{Space Telescope Science Institute, 3700 San Martin Drive, \\ 
       Baltimore, MD 21218, USA ~~ gmeylan@stsci.edu}

\begin{abstract}
Mass  determinations  are difficult  to  obtain  and still  frequently
characterised by deceptively large uncertainties.  We review below the
various  mass  estimators used  for  star  clusters  of all  ages  and
luminosities. We  highlight a few  recent results related to  (i) very
massive  old  star clusters,  (ii)  the  differences and  similarities
between  star clusters  and cores  of dwarf  elliptical  galaxies, and
(iii)  the possible  strong biases  on mass  determination  induced by
tidal effects.
\end{abstract}

%--------------------------------------------------------------------

\section{Introduction}

Open and globular clusters  are respectively located, because of their
local definition,  in the plane and  in the halo of  our Galaxy.  This
review will not make any formal distinction between these two kinds of
star clusters since (i) the above definitions apply only to our Galaxy
and (ii) we do not know  if there are any genuine differences in their
formation mechanisms.  Consequently, we may use the words ``open'' and
``globular'', but we shall essentially mean ``star clusters'', whether
they are  light or massive,  or young or  old.  This approach  is also
justified by  the fact that,  in simulations of the  Galactic globular
cluster   system,  the   dynamical  evolution   of  an   initial  mass
distribution, of either Gaussian or  power-law type, leads always to a
predicted  distribution  consistent   with  observations:  light  star
clusters  dissolve rather  quickly,  while heavy  ones survive  longer
(Baumgardt 1998, 2001).  See also Zhang  \& Fall (1999) in the case of
the populations of star clusters of the Antennae Galaxies.

Individual masses  of star  clusters are not  easy to measure.   As an
example, we  shall quote the  four different mass estimates  given for
the  giant  Galactic globular  cluster  \cent\  by Ogorodnikov  \etal\
(1976).   From  a few  stellar  radial  velocities  \Mtot\ $\simeq$  7
$\times$ $10^5  \mathcal{M}_{\odot}$, from the  same radial velocities
corrected  for the  effect of  the  internal rotation  of the  cluster
\Mtot\   $\simeq$   3   $\times$  $10^6   \mathcal{M}_{\odot}$,   from
low-quality  proper  motions measured  on  photographic plates  \Mtot\
$\simeq$ 3 $\times$ $10^7 \mathcal{M}_{\odot}$, and from gravitational
focusing  from the same  proper motions  \Mtot\ $\simeq$  2.8 $\times$
$10^8 \mathcal{M}_{\odot}$.   These four different  approaches provide
results differing by 3  orders of magnitude!  Fortunately, things have
improved since  then, but  even the best  mass determinations  of star
clusters remain rather  uncertain, typically by a factor  of 2.  It is
interesting to realize that the  age of the Universe, intuitively more
difficult to determine than the mass of a nearby star cluster, is more
accurately known since the Hubble  constant \ho\ is now estimated at a
better than 20~\% level.

%--------------------------------------------------------------------

\section{The Giant Galactic Clusters \cent\ and \tuca}

Significant  improvements  in  the   quality  and  numbers  of  radial
velocities have allowed the  determinations of reliable mass estimates
through the use  of various dynamical models.  On  the one hand, there
is  the parametric  approach.   For example,  Gunn  \& Griffin  (1979)
developed multi-mass models whose distribution functions $f$ depend on
the  stellar  energy per  unit  mass  $\varepsilon$  and the  specific
angular  momentum~$l$.  Such models  are spherical  and have  a radial
anisotropic   velocity   dispersion   ($\overline   {v_r^2}$   $\not=$
$\overline  {v_{\theta}^2}$  =  $\overline {v_{\phi}^2}$).   They  are
called King-Michie models and  associate the lowered Maxwellian of the
King model with the anisotropy factor of the Eddington models:
$$ f(\varepsilon,l) \propto 
[\exp(-2j^2\varepsilon)-\exp(-2j^2\varepsilon_t)] ~
\exp(-j^2 l^2/r_a^2) \eqno(1) 
$$
On the other hand, one never knows which of the assumptions underlying
such  a model are  adhered to  by the  real system  and which  are not
(Dejonghe \& Merritt 1992).  These  arguments suggest that it might be
profitable to interpret kinematical  data from globular clusters in an
entirely different  manner, placing much stronger demands  on the data
and making  fewer ad hoc  assumptions about the  distribution function
$f$  as well  as  the gravitational  potential  $\Phi$.  Ideally,  the
unknown  functions  should be  generated  non-parametrically from  the
data, in  an approach pioneered  by Merritt (1993, 1996).   We provide
hereafter the  results of two studies,  parametric and non-parametric,
respectively,  of  the  globular  cluster \cent,  both  studies  using
exactly  the same  observational  data, viz.,  the surface  brightness
profile and 469 stellar radial velocities.

\noindent
$\bullet$ Parametric:  a simultaneous  fit of these  radial velocities
and  of the  surface brightness  profile to  a  multi-mass King-Michie
dynamical model  provides mean estimates  of the total mass  for \cen\
equal to \Mtot\  = 5.1 \pmm\ 0.6 $\times$  \milm, with a corresponding
mean mass-to-light ratio \mlv\ = 4.1 (Meylan \etal\ 1995).

\noindent
$\bullet$ Non-parametric: the  potential and mass distribution infered
in this  method provide a total mass  for \cen\ equal to  \Mtot\ = 2.9
\pmm\  0.4 $\times$  \milm,  with a  corresponding mean  mass-to-light
ratio \mlv\ = 2.3 (Merritt \etal\ 1997).

%--------------------------------------------------------------------

\section{Universal Mass-to-Light Ratio?}

With  a  similar  parametric  approach applied  to  the  observational
constraints  obtained  for NGC~1835,  an  old  Large Magellanic  Cloud
globular  cluster, King-Michie  models  give \Mtot\  =  1.0 \pmm\  0.3
$\times$ \milm,  corresponding to a  mean mass-to-light ratio  \mlv\ =
3.4 \pmm\ 1.0 (Meylan 1988, Dubath \& Meylan 1990).

These  studies  show that  when  the  same  kind of  dynamical  models
(King-Michie) constrained  by the  same kind of  observations (surface
brightness  profile  and  central  value  of  the  projected  velocity
dispersion)  are applied  to  an old  and  bright Magellanic  globular
cluster, viz., NGC 1835, the results seem similar to those obtained in
the case  of Galactic globular  clusters.  Consequently, the  rich old
globular clusters in  the Magellanic clouds could be  quite similar in
mass and \mlv\ to the rich globular clusters in the Galaxy.
 
%--------------------------------------------------------------------

\section{The Final Word from Thousands of Stellar Space Motions}

Recently, the amount of data  related to stellar motions, viz., radial
velocities  and proper  motions,  has increased  significantly.  In  a
pioneering  ground-based study  of \cent,  van Leeuwen  \etal\ (2000)
measured the individual proper  motions of 7853 probable member stars,
from photographic  plates with epochs  ranging from 1931  through 1935
and 1978 through 1983. An  internal proper motion dispersion of 1.0 to
1.2 \masyr, equivalent to 25 to 29 \kms\ for a distance of 5.1 kpc, is
found for members near  the cluster center.  This dispersion decreases
to 0.3 \masyr, equivalent to 7.5 \kms\ in the outer regions.  The full
dynamical interpretation of these  proper motions, combined with a few
thousand radial velocities, is in preparation by the same group.

Another group,  using HST/WFPC2  images, has obtained  slightly better
measurements of proper  motions for about 15,000 stars  in the core of
\tuca, this  within a  time baseline  of only 4  years, with  3 epochs
separated  by 2  years.  See Anderson,  King,  \& Meylan  1998 for  a
progress report after the second epoch.  These data, combined with the
radial velocities of  about 5,000 stars, will provide,  as in the case
of \cent,  an insight  into the  dynamics of the  core of  \tuca, with
fundamental  by-products such  as cluster  distance and  photometry of
variable stars and binaries.

With their new proper-motion techniques and software (Anderson \& King
2000) applied to the HST/WFPC2 archive data for the first epoch and to
their own observations  for the second epoch, the  members of the same
group  will soon  have available  similar sets  of proper  motions for
about  ten among  of the  richest  and most  nearby Galactic  globular
clusters.

The  above  studies,  with  thousands  of proper  motions  and  radial
velocities constraining  dynamical models with three  integrals of the
motion as well  as non-parametric ones, will allow  a significant step
forward in our understanding of  the internal dynamics of massive star
clusters.

%--------------------------------------------------------------------

\section{Mayall~II $\equiv$~G1, a Giant Globular Cluster in M31}

Mayall~II  $\equiv$~G1  is  one  of the  brightest  globular  clusters
belonging to  M31, the Andromeda galaxy.   Observations with HST/WFPC2
provide  photometric  data  for  the $I$~vs.~$V-I$  and  $V$~vs.~$V-I$
color-magnitude  diagrams.  They reach  stars with  magnitudes fainter
than $V$  = 27  mag, with  a well populated  red horizontal  branch at
about $V$ =  25.3 mag (Meylan \etal\ 2001).   From model fitting, that
study determines  a rather high mean  metallicity of [Fe/H]  = -- 0.95
\pmm\ 0.09, somewhat similar to \tuca.  In order to determine the true
measurement errors,  Meylan \etal\ (2001) have  carried out artificial
star  experiments.  They find  a larger  spread in  $V-I$ than  can be
explained  by  the measurement  errors.   They  attribute  this to  an
intrinsic metallicity dispersion  among the stars of G1,  which may be
the consequence of  self-enrichment during the early stellar/dynamical
evolutionary phases  of this cluster.   So far, only \cent,  the giant
Galactic globular cluster, has been known to exhibit such an intrinsic
metallicity dispersion. This is, a phenomenon certainly related to the
deep  potential well of  each of  these two  star clusters,  which are
massive enough to retain the  gas expelled by the first generations of
very massive stars.

The structural  parameters of G1  are deduced from the  same HST/WFPC2
data.  Its surface brightness profile provides its core radius \rc\ =
0.14\arcs\  = 0.52 pc,  its  tidal radius  \rt\ $\simeq$ 54\arcs\  =
200~pc,  and  its  concentration  \conc\ $\simeq$ 2.5.   Such  a  high
concentration  indicates the  probable  collapse of  the  core of  G1.
KECK/HIRES observations provide the central velocity dispersion \sobs\
= 25.1~\kms, with \sigpo\ = 27.8~\kms\ once aperture corrected.

Three  estimates of the  total mass  of this  globular cluster  can be
obtained.  The King-model mass is  \Mass$_K$ = 15 $\times$ \milm\ with
\mlv\  $\simeq$  7.5, and  the  Virial  mass  is \Mass$_{Vir}$  =  7.3
$\times$ \milm\ with \mlv\ $\simeq$ 3.6.  The King-Michie model fitted
simultaneously  to  the surface  brightness  profile  and the  central
velocity  dispersion  value   provides  mass  estimates  ranging  from
\Mass$_{KM}$ = 14 $\times$ \milm\ to 17 $\times$ \milm\ (Meylan \etal\
2001).

\begin{table}
\caption{Three mass determinations for Mayall~II $\equiv$~G1 and \cen}
\begin{center}
\vskip 6pt
\begin{tabular}{ccc}
\tableline
\tableline
 Mass & Mayall~II   & \cent    \\ 
      &   [\milm]   & [\milm]  \\ 
\tableline
 King        &  15     & 4.3  \\ 
 Virial      & 7.3     & 2.9  \\ 
 King-Michie & 13-18   & 5.1  \\ 
\tableline
\end{tabular}
\end{center}
\end{table}

The  spread between  the  three mass  determination  values listed  in
Table~1  give a  better idea  than their  individual  formal (smaller)
errors about  their true uncertainties.   The masses of  both clusters
are known to about a factor of two.  Although not very precise, all of
these mass estimates make G1 more  than twice as massive as \cent, the
most massive Galactic  globular cluster.  G1 is unique  in M31 because
of its projected  location 40~kpc away from the  center of the galaxy,
but there  are at least three  other bright globular  clusters in this
galaxy  which have  velocity dispersions  \sobs\ larger  than 20~\kms,
implying rather large masses.

%--------------------------------------------------------------------

\section{On the Origin of the Most Massive Globular Clusters}

Such large masses are related  to the metallicity spread the origin of
which  is still  unknown.  It  may  come either  (i) from  metallicity
self-enrichment in  a massive  globular cluster, (ii)  from primordial
metallicity inhomogeneity in a binary proto-cluster cloud, followed by
early merger, or (iii) from the fact that the present globular cluster
is  merely the  remaining core  of a  previously larger  entity, e.g.,
originally a dwarf galaxy subsequently pruned by dynamical evolution.

Although \cent\ is  the best studied globular cluster,  because of its
size and relative proximity  ($\sim$ 5.1~kpc), many conundrums remain:
viz., (i) the metallicity spread among stars (Freeman \& Norris 1981),
(ii)  its double Main  Sequence (Anderson  1997), (iii)  the different
kinematics between  metal-rich and  -poor stars (Norris  \etal\ 1997),
and (iv) a correlation between metallicity and age, implying that this
cluster  enriched itself  over  a timescale  $\sim$  3~Gyr (Hughes  \&
Wallerstein 2000 and Hilker \& Richtler 2000).

\begin{table}
\caption{Parameters  for   the  massive  globular   cluster  Mayall~II
$\equiv$~G1  and the nucleus  of the  dwarf elliptical  galaxy NGC~205
(see Meylan \etal\ 2001)}
\begin{center}
\vskip 6pt
\begin{tabular}{ccccc}
\tableline
\tableline
 Parameters && Mayall~II   && NGC 205    \\
\tableline
 \sigpo &   &  27.8  \kms & &  30    \kms  \\
 \Mv    &   &--10.94 mag  & & --9.6  mag \\
 \rc    &   &   0.52 pc   & &   0.35 pc  \\
 \muov  &   &  13.47 mag/arcs$^{2}$ & & 12.84 mag/arcs$^{2}$ \\
\tableline
\end{tabular}
\end{center}
\end{table}

\begin{figure} 
\centerline{\vbox
{
\psfig{figure=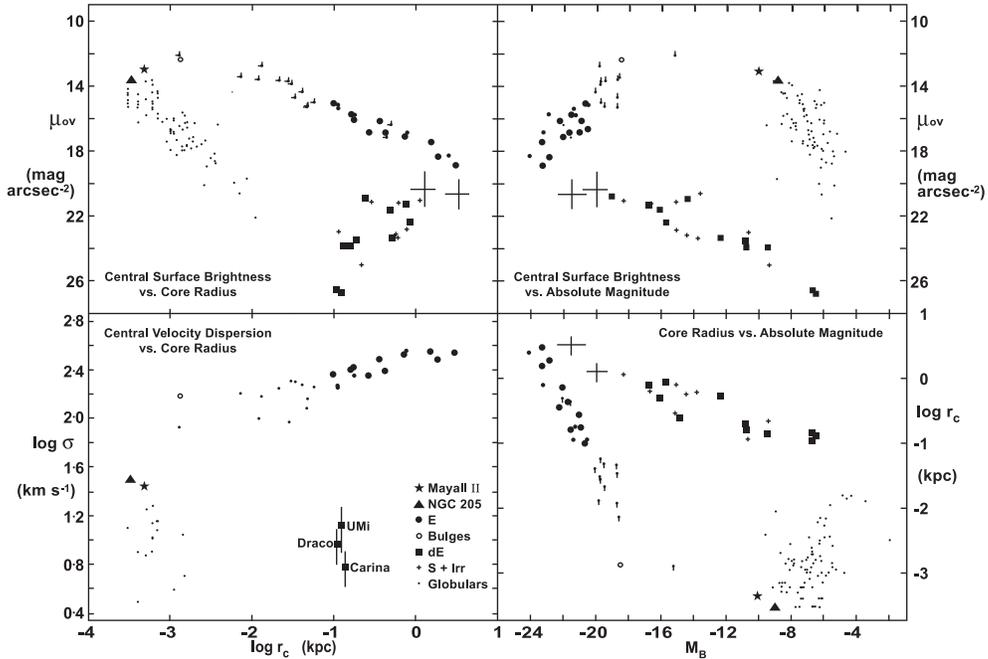,width=13cm,angle=0}
}}
\caption{ Reproduction of Fig.~3 from Kormendy (1985) in which we have
added  the  positions,  using  the  four  parameters  in  Table~2,  of
Mayall~II   $\equiv$~G1  (star)   and  of   the  nucleus   of  NGC~205
(triangle).  These fall  in all  four  panels right  on the  sequences
defined  by  globular clusters  and  always  away  from the  sequences
defined by ellipticals, bulges, and dwarf ellipticals.}
\end{figure}

Let us  consider (see  Table~2 and Meylan  \etal\ 2001)  the following
four  parameters  relative  to  G1: the  central  velocity  dispersion
\sigpo\ = 28~\kms, the integrated  absolute visual magnitude \Mv\ = --
10.94 mag,  the core  radius \rc\ =  0.52~pc, and the  central surface
brightness \muov\  = 13.47 mag~arcsec$^{-2}$.  The positions  of G1 in
the  different diagrams defined  by Kormendy  (1985), using  the above
four parameters,  always put  it on the  sequence defined  by globular
clusters,  and definitely  away from  the other  sequences  defined by
elliptical galaxies,  bulges, and dwarf  spheroidal galaxies (Fig.~1).
The same is true for \cent\ (Meylan \etal\ 2001).

Little is known about the  positions, in these diagrams, of the nuclei
of nucleated dwarf elliptical galaxies, which could be the progenitors
of the  most massive, if  not all, globular clusters  (Zinnecker 1988,
Freeman  1993).  The  above four  parameters  are known  only for  the
nucleus of one  dwarf elliptical, viz., NGC~205, and  their values put
this object, in Kormendy's diagram, close to G1, right on the sequence
of globular clusters  (see Table~2 and Fig.~1).  This  result does not
prove by  itself that  all massive globular  clusters are  the remnant
cores of nucleated dwarf galaxies.

At  the moment,  only  the anti-correlation  of  metallicity with  age
recently observed in \cent\ suggests that this cluster enriched itself
over a time scale of about 3~Gyr (Hughes \& Wallerstein 2000 and Hilker
\& Richtler  2000).  This  contradicts the general  idea that  all the
stars in  a globular cluster are  coeval, and may favor  the origin of
\cent\ as  being the  remaining core  of a larger  entity, e.g.,  of a
former nucleated  dwarf elliptical galaxy.   In any case, by  the mere
fact that  their large masses imply complicated  stellar and dynamical
evolutions,  the very massive  globular clusters  may blur  the former
clear (or  simplistic) difference between globular  clusters and dwarf
galaxies.

%--------------------------------------------------------------------

\section{Mass Estimates of Young Star Clusters}

So far,  we have  discussed only mass  estimates related to  old, rich
star clusters.   HST has triggered  numerous studies of  young, bright
starburst clusters  which may be quite massive.   See, e.g., Holtzmann
\etal\  (1992) in  the case  of  NGC~1275 clusters,  and Schweizer  \&
Seitzer (1993) and Whitmore \etal\ (1993) in the case NGC~7252.  Conti
\& Vacca (1994) use HST/FOC UV  imaging of various bright knots in the
nuclear  starburst  region of  the  Wolf-Rayet  galaxy  H2 2-10.   The
luminosities of the knots are compared to those predicted from stellar
population synthesis models for ages  between 1 and 10~Myr, from which
they  estimate  the masses  of  the knots  to  be  between 10$^5$  and
10$^6$~\msun.  In a  more direct way, B\"oker \etal\  (1999) measure a
velocity  dispersion \sobs\  = 33  \pmm\ 3~\kms\  in the  nuclear star
cluster of the  face-on giant Scd spiral galaxy  IC~342 and deduce, by
fitting the central surface  brightness profile, a cluster mass \Mtot\
=  6 $\times$ \milm.   They infer  a best-fitting  cluster age  in the
range 10 - 60 $\times$ \mily.

It is worth emphasizing that mass estimates of young star clusters are
more  difficult to  obtain  and  more uncertain  than  those of  older
clusters. This is due to  the   frequent  lack  of  velocity  dispersion
measurement and because of the large uncertainties inherent in the use
of stellar population synthesis models for young stellar systems whose
luminosities evolve quickly (see, e.g., Bruzual 2001).

%--------------------------------------------------------------------

\section{The Slow Destruction of Star Clusters}

Numerical works  have demonstrated that  continual two-body relaxation
within globular  clusters combined with weak  tidal encounters between
globular clusters and the Galactic disk and/or bulge will lead, around
each globular  cluster, to the development  of both a  halo of unbound
stars and  tidal tails.   All clusters observed,  which do  not suffer
from  strong  observational  biases,  present tidal  tails  (Grillmair
\etal\ 1995,  Leon \etal\ 2000).  These tidal  tails exhibit projected
directions preferentially  aligned with  the cluster orbit  or towards
the  Galactic  center,  betraying  their  recent  dynamical  evolution
through  disk  and/or bulge  shocking.   See  also Odenkirchen  \etal\
(2001).

Recent  theoretical work  corroborate these  observations.   In N-body
simulations  of globular  clusters  moving in  the Galactic  potential
well, Combes \etal\ (1999) observe that once the particles (stars) are
unbound, they  slowly drift along  the globular cluster path  and form
two huge  tidal tails.   A cluster is  always surrounded by  two giant
tidal tails and debris, in  permanence along its orbit.  The length of
these  tidal  tails is  of  the  order 5  tidal  radii  or more.   The
orientation of  these tidal  tails is the  signature of the  last disk
crossing and can constrain strongly the cluster orbit and the Galactic
model.  Each  disk/bulge crossing may extract  up to about  1\% of the
total  mass of  the star  cluster, leading  slowly but  surely  to its
complete evaporation.   The lighter the  cluster and the  stronger the
tidal shocks, the faster the destruction process.

%--------------------------------------------------------------------

\section{Do Globular Clusters Have Dark Halo?}

Are  globular  clusters  the  most massive  stellar  systems  without
non-baryonic dark matter?  Since  both the velocity dispersion profile
and the rotation  curve of \cent\ decrease towards  the outer parts of
this cluster, we  may conclude that there is  no dynamical evidence of
any massive halo made of non-baryonic dark matter.

In  order to  investigate further  this  point among  the most  poorly
studied  Galactic  clusters,  C\^ot\'e  \etal\  (2001)  have  obtained
KECK/HIRES high-accuracy  radial velocities of  about 20 stars  in six
outer-halo globular clusters, located between  20 and 100 kpc from the
center of our Galaxy.  The velocity dispersions range between \sobs\ =
1 and 5  \kms, with corresponding \mlv\ values between  1 and 4.  This
is  exactly what  is expected  from the  nearby well  studied globular
clusters such as \cent\ and \tuca.

However there  is one single,  although conspicuous, exception  in the
sample  of   C\^ot\'e  \etal\  (2001):  Pal~13   exhibits  a  velocity
dispersion \sobs\  = 2.6 \pmm\ 0.3  \kms.  With its  low and uncertain
total luminosity, Pal~13 has a corresponding \mlv\ in the range 10 $<$
\mlv\  $<$  40.   This  is   quite  unique  for  a  globular  cluster.
Simulations  show   that  such  a  result  would   be  mimicked,  with
difficulties, only by an uncomfortably large fraction of binary stars.
Binaries exist in globular clusters,  but not in large quantities.  We
should apply Ockham's razor before  invoking the presence of a massive
dark  halo.  Such  a high  \mlv\  value could  be an  indication of  a
cluster in  the late phases of  its dynamical evolution,  when a large
fraction of  its total mass is  made of white dwarfs,  as predicted by
Vesperini \& Heggie (1997).  It  could also be explained by a velocity
dispersion  inflated because  of  lack of  Virial equilibrium:  Pal~13
seems  to be  in the  advanced stages  of tidal  disruption,  a status
supported by  the recent observations of  extra-tidal extension around
this cluster (Siegel \etal\ 2001 and C\^ot\'e \etal\ 2001).  These two
explanations do not exclude each other.

%--------------------------------------------------------------------

\begin{figure} 
\centerline{\vbox
{
\psfig{figure=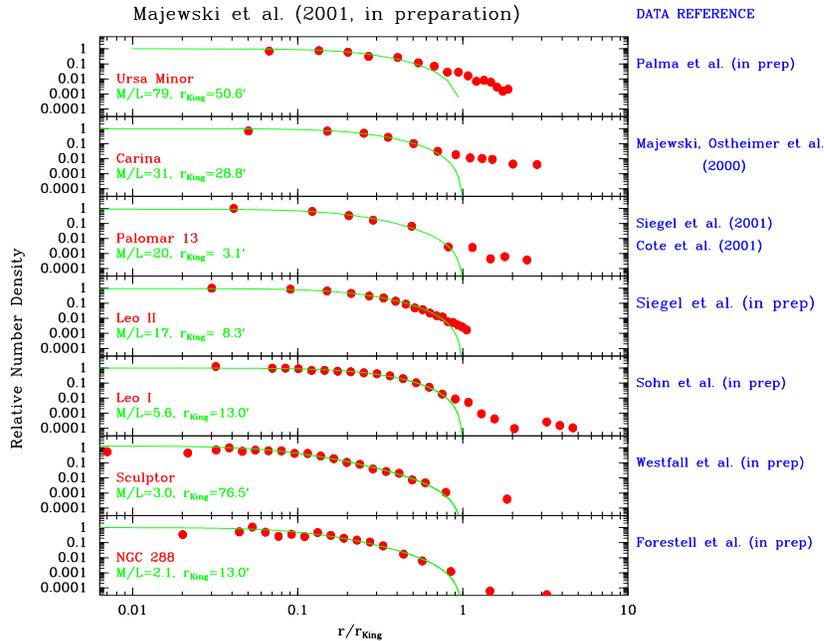,width=13cm,angle=270}
}}
\caption{From  bottom to  top,  a sequence  of increasingly  disturbed
surface  brightness  profiles (two  globular  clusters  and five  dSph
galaxies) corresponding to increasing \mlv\ values.}
\end{figure}

\section{Is Tidal Disruption an Ubiquitous Phenomenon?}

Leon \etal\  (2000) show that  any Galactic globular  cluster observed
which  does  not  suffer   from  strong  observational  biases  (e.g.,
intervening  absorption by  Galactic cirrus  along  the line-of-sight)
displays a  pair of  tidal tails.  The  tidal disturbance may  also be
unveiled  through its  effect on  the surface  brightness  profile. An
isolated  cluster  suffering  no   tidal  shock  will  have  a  normal
King-model profile  characterised by a core surrounded  by an envelope
with  a steep  profile.  Any  tidally-perturbed cluster  will  have an
envelope  profile departing from  the King-model  profile as  if there
were a very strong background of stars: the stronger the tidal effect,
the higher the level of the background (i.e., the higher the departure
from King-model profile), which is  made of stars escaping the cluster
gravitational attraction.

When observing a  tidally perturbed globular cluster, there  is a high
probability of  measuring radial  velocities of escaping  stars, which
are no more in Virial  equilibrium with the cluster.  Such stars would
inflate  the measured  velocity  dispersion.  The  stronger the  tidal
disturbance,  the higher  the  inflated velocity  dispersion, and  the
larger the mass-to-light  ratio.  This is exactly what  is observed in
Fig.~2,  from  Majewksi  \etal\  (2001), which  displays  the  surface
brightness profiles of two globular clusters (NGC~288 with \mlv\ = 2.1
and  Pal~13 \mlv\  $\sim$ 20)  and five  dSph galaxies  (Sculptor with
\mlv\ =  3.0, Leo~I with \mlv\ =  5.6, Leo~II with \mlv\  = 17, Carina
with \mlv\  = 31, and  Ursa Major with  \mlv\ = 79).  From  the bottom
panel (NGC~288) to  the top one (Ursa Major) we  observe a sequence of
increasingly  disturbed profiles,  with higher  and  higher departures
from the King  models, with the most disturbed  profile being for Ursa
Major.   Interestingly,   this  sequence  of   increasingly  disturbed
profiles correspond to a  sequence of increasing mass-to-light ratios,
as if the  \mlv\ values were a direct measure of  the intensity of the
tidal  shocks.  Is  the  evidence  for dark  matter  in dSph  galaxies
evaporating with their stars?

%--------------------------------------------------------------------

\acknowledgments

It  is  a   pleasure  to  thank  Steve  Majewksi   (Virginia)  and  my
collaborators  Patrick   C\^ot\'e  (Rutgers)  and   George  Djorgovski
(Caltech)  for   allowing  me  to   present  results  in   advance  of
publication.  I also thanks Jennifer Lotz (JHU), Roeland van der Marel
(STScI),  and Brad  Whitmore (STScI)  for interesting  discussions and
information about NGC~205.

%--------------------------------------------------------------------

%--------------------------------------------------------------------

\section*{Discussion}

\noindent  {\it Armandroff:\,}  You  raised the  possibility of  tidal
disruption  leading   to  the   large  velocity  dispersion   in  Ursa
Minor. Models for the disruption  of dSphs (Piatek \& Pryor, 1995, AJ,
109,  1071;  Oh  \etal,  1995,  ApJ, 442,  142)  predict  a  kinematic
signature that  resembles a  rotation curve, as  opposed to  simply an
increased dispersion.  Large samples of radial velocities in Draco and
Ursa  Minor have  not shown  evidence for  the kinematic  signature of
disruption.   Do you  have  evidence  for Ursa  Minor  or other  dSphs
matching the modeling? \\

\noindent {\it  Meylan:\,} The two papers you  mention present results
of  N-body simulations  which reach  conclusions different  from those
obtained from other, more recent N-body simulations (Kroupa, 1997, New
Astron, 2,  139; Klessen  \& Kroupa, 1998,  ApJ, 498,  143).  Kroupa's
work shows that  dSphs are obtained under the  extreme hypothesis that
dSph   progenitors  are  not   dominated  by   dark  matter   but  are
significantly shaped by tides through many periastron passages; Piatek
\&  Pryor  simulated  only  one  passage,  and  Kroupa's  results  are
completely consistent with their findings for one passage only.  There
are now in Draco photometric observations of stars beyond its measured
tidal  boundary  (Piatek  \etal,  2001  AJ, 121,  841).   It  will  be
essential to obtain, for a few of these dSphs, samples of thousands of
stellar  radial  velocities.   At   the  moment,  I  simply  find  the
correlation between tidal disturbance intensity and \ml\ values rather
intriguing. This definitely calls for more studies. \\

\noindent {\it Da Costa:\,} Comment: Integrated spectrum of nucleus of
NGC~205 is  A-type, so its  stellar population is very  different from
globular clusters  like G1. Question: What  is the core  radius of the
NGC~205 nucleus and does if fit  a King model? (different from say M32
with its power-law cusp). \\

\noindent  {\it  Meylan:\,}  About   the  comment:  Yes,  the  stellar
populations in the core of NGC~205 are younger than those in G1.  This
means that the position, in the panels defined by Kormendy (see Fig.~1
above), of a  dynamical system older than about  1~Gyr does not depend
strongly on  its age. Answer to  the question: The core  radius of the
nucleus NGC~205  is 100 mas  = 0.35 pc  (Jones \etal, 1996,  ApJ, 466,
742), similar to a very dense globular cluster easily fitted by a King
model.  The  star cluster in the  center of NGC~205  is very different
from  the nucleus  of M32,  which has,  e.g., a  much  larger velocity
dispersion \sobs\ = 150 \kms  (Joseph \etal, 2001, ApJ, 550, 668). See
also Meylan \etal\ (2001). \\

\noindent  {\it Grillmair:\,}  Two  comments: 1)  The  onset of  tidal
effects  (e.g., break  from normal  profile to  power-law  tidal tail,
velocity dispersion increase) can  depend significantly on the orbital
phase of the cluster (strongest effect at apogalacticon). 2) Ben Moore
(1996, ApJ, 461, L13) put an  upper limit of \ml\ of NGC~7089 $\equiv$
M2, based on existence of tidal tails. \\

\noindent {\it Wenderoth:\,} Around 15 years ago, the idea that \cent\
is  a merger  was  proposed. Since  then,  is there  more evidence  to
support this idea or to reject it? \\

\noindent {\it Meylan:\,} There was  a paper by Icke \& Alcaino, 1988,
A\&A,  204, 115,  suggesting  that \cent\  could  be the  result of  a
merger, this in  order to explain (i) its  spread in metallicity, (ii)
its strong  flattening and  (iii) its large  mass.  There is  now more
evidence about  the intrinsic complexity of this  globular cluster, as
mentioned  above in  Section~6, although  none of  these observational
facts supports the merger scenario over the other two alternatives. If
merger there was, it must have been  very early in the life of the two
merging proto-cluster clouds, since  the mere encounter of two current
globular clusters  would not  induce a merger,  except in the  case of
orbital-parameter adjustment with vanishingly small probabilities. \\

%--------------------------------------------------------------------


\begin{references}
\reference Anderson J., 1997, Ph.D. thesis, 
           University of California, Berkeley
\reference Anderson J., King I.R., Meylan G., 1998, BAAS, 30, 1347
\reference Anderson J., King I.R., 2000, PASP, 112, 1360
\reference Baumgardt H., 1998, A\&A, 330, 480
\reference Baumgardt H., 2001, in ASP Conf. Ser.  Vol.  ???, Modes of
           Star  Formation  and   the  Origin  of  Field  Populations,
           ed. E.K. Grebel and W. Brandner 
           (San Francisco: ASP), in press
\reference B\"oker T., van der Marel R.P., Vacca W.D., 
           1999, AJ, 118, 831
\reference Bruzual G.A., 2001, in 
           XI Canary Islands Winter School of Astrophysics, 
           Galaxies at High Redshift, 
           ed. I. P\'erez-Fournon and F. S\'anchez, 
           (Cambridge: Cambridge Contemporary Astrophysics), in press
\reference Combes F., Leon S., Meylan G., 1999, A\&A, 352, 149
\reference Conti P.S., Vacca W.D., 1994, ApJ, 423, L97
\reference C\^ot\'e P., Djorgovski S.G., Meylan G., 
           Castro S., McCarthy J.K., 2001, AJ, submitted
\reference Dejonghe H., Merritt D., 1992, ApJ, 391, 531
\reference Dubath P., Meylan G., Mayor M., Magain P.,
           1990, A\&A, 239, 142
\reference Freeman K.C., 1993, in IAU Symp. 153, Galactic Bulges, 
           ed. H. Dejonghe \& H.J. Habing (Dordrecht: Kluwer), 263
\reference Freeman K.C., Norris J.E., 1981,   ARA\&A, 19, 319
\reference Grillmair C. J., Freeman K. C., Irwin M., 1995, AJ, 109, 2553 
\reference Gunn J.E., Griffin R.F., 1979, AJ, 84, 752
\reference Hilker M., Richtler T., 2000, A\&A, 362, 895
\reference Holtzmann J.A., \etal\, 1992, AJ, 103, 691
\reference Hughes J., Wallerstein G., 2000, AJ, 119, 1225
\reference Kormendy J., 1985, ApJ, 295, 73
\reference Leon S., Meylan G., Combes F., 2000, A\&A, 359, 907
\reference Majewksi S.R., \etal, 2001, in preparation
\reference Meylan G., 1988, ApJ, 331, 718
\reference Meylan G., Heggie D.C., 1997, A\&AR, 8, 1-143
\reference Meylan G., Mayor M., Duquennoy A., Dubath P., 1995, 
           A\&A, 303, 761
\reference Meylan G., Sarajedini A., Jablonka P., 
           \etal, 2001, AJ, in press
\reference Merritt D., 1993, ApJ, 413, 79
\reference Merritt D., 1996, AJ, 112, 1085
\reference Merritt D., Meylan G., Mayor M., 1997, AJ, 114, 1074
\reference Norris J.E., Freeman K.C., Mayor M., Seitzer P., 1997, 
           ApJ, 487, L187
\reference Odenkirchen M., Grebel E.K., \etal, 2001, ApJ, 548, L165
\reference Ogorodnikov K.F., Nezhinskii E.M., Osipkov L.P., 1976,
           Sov. Astr. Lett. 2, 57
\reference Schweizer F., Seitzer P., 1993, ApJ, 417, L29
\reference Siegel M.H., Majewski S.R., Cudworth K.M., Takamiya M., 
           2001, AJ, 121, 935
\reference van Leeuwen F., Le Poole R.S., Reijns R.A., 
           Freeman K.C., de Zeeuw P.T., 2000, A\&A, 360, 472
\reference Vesperini E., Heggie D.C., 1997, MNRAS, 289, 898
\reference Whitmore B.C., Schweizer F., Leitherer C., 
           Borne K., Robert C., 1993, AJ, 106, 1354
\reference Zhang Q., Fall S.M., 1999, ApJ, 527, L81
\reference Zinnecker H., Keable C.J., Dunlop J.S., 
           Cannon R.D., Griffiths W.K., 1988, in IAU Symp. 126, 
           {\sl Globular Cluster Systems in Galaxies}, 
           ed. J.E. Grindlay, A.G.D. Philip, 
           (Dordrecht: Kluwer), 603
\end{references}
\end{document}